\documentclass[12pt]{article}
\usepackage{graphicx}
\usepackage{floatrow}
\newfloatcommand{capbtabbox}{table}[][\FBwidth]
\usepackage{amsmath,amssymb,amsthm}
\usepackage{latexsym,graphicx,color,subfigure,slashed,mathrsfs,verbatim,bbm,wasysym, pstricks,epsfig,url, hyperref}
\usepackage{indentfirst}
\usepackage[sort&compress,numbers]{natbib}
\usepackage{epstopdf}
\usepackage{nicefrac}
\usepackage{authblk}
\usepackage{ dsfont }
\usepackage[font=footnotesize]{caption}
\usepackage{tikz}
\usetikzlibrary{shapes,snakes}
\usetikzlibrary{positioning}
\usetikzlibrary{calc}
\usepackage{color,soul}
\usepackage[toc]{appendix}
\numberwithin{equation}{section}

\def\beq{\begin{equation}}
\def\eeq{\end{equation}}
\def\bea{\begin{eqnarray}}
\def\eea{\end{eqnarray}}
\def\bit{\begin{itemize}}
\def\eit{\end{itemize}}
\def\mcal{\mathcal}
\def\G{\mcal{G}}

\begin{document}

\title{Signatures of an $S_3$-Orbifold Higgs Model}
\author{Keith Thrasher\thanks{rkthrasher@email.wm.edu}}
\affil{\footnotesize{High Energy Theory Group, College of William and Mary, Williamsburg, Virginia 23187, U.S.A.}}

\maketitle

\begin{abstract}
The lack of new physics at the LHC has sparked renewed interest in theories of neutral naturalness, in 
which the Standard Model partners required for naturalness carry no Standard Model charge.
The Twin Higgs was the first of these theories
to be introduced, but recent work has demonstrated it is only an isolated example in a large class of 
``orbifold Higgs" models. In this work we study an orbifold Higgs model resulting from the orbifold projection
by the non-abelian group $S_3$. A model with multiple sectors uncharged under the SM emerges. Constraints
are placed on the model from Higgs phenomenology and the prospects of finding evidence at colliders are discussed.
\end{abstract}

\newpage
\section{Introduction}
%======================================================%
	
	The discovery of the Higgs boson\cite{Aad:2012tfa,Chatrchyan:2012ufa} has 
provided us with the last piece needed to complete the Standard Model (SM). Due to radiative 
corrections to the Higgs mass term, the SM requires an extreme fine tuning in order to keep 
the weak scale much smaller than the Planck scale. With the belief that such a tuning in nature 
is unnatural, many solutions have been proposed to eliminate the large quadratic corrections to the Higgs mass,
 thus eliminating the hierarchy problem. Supersymmetry and compositeness are prime examples of such theories,
but current null search results for SM partners are now forcing many of these
models into finely tuned territory. The fine tuning that is necessary to create a hierarchy between the weak scale
and the scale which new physics appears is called the little-hierarchy problem\cite{Barbieri:2000gf}.
		
	The Twin Higgs\cite{Chacko:2005pe,Chacko:2005un} is a solution to the little-hierarchy problem where the SM
Higgs is played by the role of a pseudo-Goldstone boson. The SM fields are joined by a set of partners
called ``twin" states. These SM partners differ in comparison to those in supersymmetry in that they carry no SM
charge. This would make current searches for partner states to the SM especially challenging and may
explain current null search results at the LHC. A discrete $\mathds{Z}_2$ symmetry that interchanges the 
SM fields with the twin states then ensures gauge, Yukawa, and scalar self interactions must be equivalent 
in the SM and twin sectors. This protects the pseudo-Goldstone Higgs against the quadratic corrections the 
Higgs mass term receives in the SM. Typically for cutoff scales
$\Lambda \sim 5-10\mbox{ TeV}$ these models do not suffer from any major fine tuning. At higher scales
a stronger mechanism such as compositeness or SUSY may keep the weak scale natural
to the Planck scale as demonstrated in UV completions of the Twin Higgs\cite{Batra:2008jy,Barbieri:2015lqa,
Low:2015nqa,Badziak:2017syq,Katz:2016wtw,Geller:2014kta,Craig:2013fga,Falkowski:2006qq,Chang:2006ra}.

	Other theories of neutral naturalness have since been introduced\cite{Burdman:2006tz,Cai:2008au,Chacko:2005vw,Batell:2015aha,Arkani-Hamed:2016rle}, including
recent work which has demonstrated that the Twin Higgs is only the simplest example in a large class
of orbifold Higgs models\cite{Craig:2014aea,Craig:2014roa}. In orbifold Higgs models, the Higgs is protected by an accidental
symmetry resulting from an orbifold reduction of a larger symmetry via some discrete group. These models 
also generically give rise to states that are uncharged under the SM. The orbifold interpretation also lends
itself nicely in creating UV complete models as geometric orbifolds of some higher dimensional space.

	In this paper we explore one of these orbifold Higgs models arising from a non-abelian
orbifold pattern, namely $S_3$. Like the Twin Higgs this produces hidden sectors, one SM-like
in structure and another exotic sector with an $SU(6)$ color group, $SU(4)$ weak isospin group, and an $SU(2)$
flavor symmetry among the Higgs and top partners. Though the model has been specified in the
original orbifold Higgs papers, the details of the experimental signatures have yet to be 
carried out. In this paper we explore the phenomenology of the $125 \mbox{ GeV}$ SM-like Higgs generated by the
model and compare results to the signatures predicted in the Twin Higgs.

	In the next section we will review the features of the Twin Higgs. Following this the
formalism behind field theory orbifolds will be given as a necessity to understand how
orbifold Higgs models are constructed. The $S_3$-orbifold Higgs will then be presented and 
we'll demonstrate how a natural SM-like Higgs emerges from the model. Section \ref{sec:pheno}
will analyze some of the phenomenology and compare the results to the Twin Higgs and 
section \ref{sec:c&o} will contain our conclusions.

\section{Twin Higgs Review}
%======================================================%
	We will now take a moment to review the Mirror Twin 
Higgs\cite{Chacko:2005pe}. We begin with a complex 
scalar, $H$, which transforms as a fundamental of a global 
$SU(4)$ symmetry. The scalar potential is given by,
\beq
V = -m^2 |H|^2 + \lambda |H|^4
\eeq
where $m^2>0$. $H$ picks up a vacuum expectation 
value (vev), $|\langle H \rangle| \equiv \frac{f}{\sqrt{2}}$, and the 
global symmetry is broken to $SU(4) \rightarrow SU(3) $
yielding 7 massless Goldstone bosons. 

	We now explicitly break the global $SU(4)$ by gauging the subgroup 
$SU(2)_A \times SU(2)_B \subset SU(4)$ such that H transforms 
as $H^T=(H_A \,\, H_B)$. After gauging this symmetry the global $SU(4)$ 
symmetry is still an accidental symmetry of the tree level potential. In general, 
radiative corrections to the potential will not be invariant under the accidental
 $SU(4)$. For instance the Higgs gauge interactions generate terms such as
\beq
\Delta V \sim \frac{9 \Lambda^2}{16 \pi^2} \left(g_A^2 |H_A|^2 + g_B^2 |H_B|^2  \right),
\eeq
where we have used a uniform hard cutoff to regulate the integrals. This 
introduces mass terms for the Goldstones that are quadratically sensitive
 to the cutoff. We can eliminate this by introducing a discrete $\mathbb{Z}_2$ 
 symmetry, dubbed twin-parity. This symmetry exchanges the gauge fields and 
 $H_A \leftrightarrow H_B$ which enforces that the gauge couplings are
 equal, $g \equiv g_A=g_B$. Now, 
\beq
	\Delta V \sim \frac{9 g^2 \Lambda^2}{16 \pi^2} \left( |H_A|^2 +  |H_B|^2  \right) 
	= \frac{9 g^2 \Lambda^2}{16 \pi^2} \left( |H|^2  \right) 
\eeq
which is an $SU(4)$ invariant. Thus the quadratic divergences do not contribute 
to the masses of the Goldstone bosons. From here we can create twin copies of the 
fermions and gluons and extend twin parity to the twin gluons and fermions. 
This will eliminate the quadratic divergences due to the Yukawa interactions. 
The Higgs mass term and quartic interactions arise from $SU(4)$ breaking terms 
stemming from the one-loop effective potential. 

	Without additional soft terms added to the potential neither sector is suited
to be identified with the SM sector as the Higgs would be equally aligned with both 
$A$ and $B$ sectors. This would lead to a $1/\sqrt{2}$ suppression in the couplings
of the Higgs to the SM which isn't consistent with experiment. To identify the $A$-sector with the SM
we can add $V_{soft} = \mu |H_A|^2$ to the potential which softly break twin parity. Tuning
the soft term, $\rho$, against the $SU(4)$ breaking order parameter, $f$, will suppress the $A$-sector
Higgs couplings to $B$-sector states by $\sin(v/f)$ where v is the vev of the SM Higgs. For $v \ll f$
this provides a phenomenologically viable scenario where the SM is associated with the $A$-sector.
We will see in the following sections how the Twin Higgs paradigm can be generalized
by way of the orbifold Higgs and how the quadratic divergences are eliminated (or at least suppressed) 
in general orbifold Higgs theories.

\section{Building an Orbifold Higgs Model}
%======================================================%

In this section we will briefly review field theory orbifolds which will be vital to 
understanding orbifold Higgs models. For a more detailed approach of what 
follows we refer the reader to ref. \cite{Craig:2014roa, Schmaltz:1998bg}. 

\subsection{Field Theory Orbifolds}
%======================================================%

	Let's begin with some initial field theory, called the parent theory, which has some global 
or gauge symmetry, $G$. To orbifold the parent symmetry by some discrete group, $\G$, we must
study the action of $\G$ on $G$. This requires that we first embed $\G$ into the parent theory 
which we will do through the regular representation embedding. The fields in the parent theory 
that are left invariant under the action under $\G$ will be those that comprise the daughter theory 
and all other states are projected out. 

	As an example, consider a parent theory consisting of a scalar, $H$, which transforms as 
a bifundamental of a gauged $SU(\Gamma N)$  and global $SU(\Gamma F)$, where $F,N\in \mathds{N}$, as shown in 
Figure \ref{tab:sp} . We'll then take our discrete group, $\G$, to be of order, $|\G| = \Gamma$.  
\begin{figure}[h]
\begin{tabular}{c|cc}
        		&	$SU(\Gamma N)$	&	$SU(\Gamma F)$ 	\\ \hline
        $H$ 	& 	$\square$ 		& 	$ \overline\square$
\caption{{\small{Transformation properties of the scalar field $H$ in the parent theory. }}}
\label{tab:sp}
\end{tabular}
\end{figure}
We now need to determine the orbifold of the parent theory by $ \G$. First, we 
express $\G$ in the regular representation which has the following well known decomposition,
\beq
\begin{array}{cc}
\gamma^s_R = \bigoplus\limits_{\alpha = 1}^{n_{\G}} \mathds{1}_{d_\alpha} \otimes r^s_\alpha &\hspace{5 mm} s \in 1 ... \Gamma.
\end{array}
\eeq
Here, $s$ labels the elements of the group, $r_\alpha$ denotes the irreducible representations of $\mcal{G}$ 
with relative dimension $d_\alpha$, and $\alpha$ sums over the $n_\G$ irreducible representations. To embed 
$\G$ into $SU(N \Gamma)$ we take the direct product of the N-dimensional identity and regular representation yielding,
\beq
\gamma^s_{N} \equiv \mathds{1}_N \otimes \gamma^s_R = \bigoplus\limits_\alpha \mathds{1}_{Nd_\alpha} \otimes r_\alpha^s.
\eeq

We can now study the transformation properties of the fields in the parent theory under action of $\gamma_N$ and project out all fields not invariant under the action. For fields transforming in the adjoint representation, the invariant states are those satisfying,
\beq
A = \gamma^s_{N} \, A\, \left(\gamma^s_{N} \right)^\dagger
\eeq
for all $s\in \{1 ... \Gamma \}$.The orbifold of $ SU(\Gamma N)$ by $\mcal{G}$ reduces the symmetry to a direct product of smaller symmetry groups in the daughter theory, namely
\beq
SU(\Gamma N) \longrightarrow \left( \prod\limits_{\alpha = 1}^{n_{\mcal{G}}} SU(d_\alpha N) \right) \otimes \left(U(1) \right)^{n_{\mcal{G}}-1}.
\eeq

To find the invariant components of fields transforming in the fundamental representation it's convenient to construct projection operators. For the field $H$ transforming as a bifundamental of $SU(\Gamma N)\times SU(\Gamma F)$ the projection operator takes the form, 
\beq
P_R = \frac{1}{\Gamma} \sum\limits^\Gamma_{s=1} \gamma_N^s \otimes (\gamma_F^s)^{*} ,
\eeq
where $P_R$ acts on the left of $H$. This procedure will in general leave us with a daughter theory with non-canonically normalized kinetic terms with rescaling related to the dimension of the representation, $d_\alpha$. Requiring normalized kinetic terms in the orbifold daughter theory induces a rescaling of the interactions of the daughter theory. Scalar masses, $m$, and double-trace quartic interactions $\lambda$ in the parent theory don't get rescaled in the daughter, gauge couplings, $g$, and yukawas, $y$, of the parent get rescaled by $1/\sqrt{d_\alpha}$, and single trace quartics get rescaled by $1/d_\alpha$.

\subsection{Orbifold Higgs}
%======================================================%

\begin{figure}[h]
\begin{floatrow}
\ffigbox{%
\includegraphics[scale = .18]{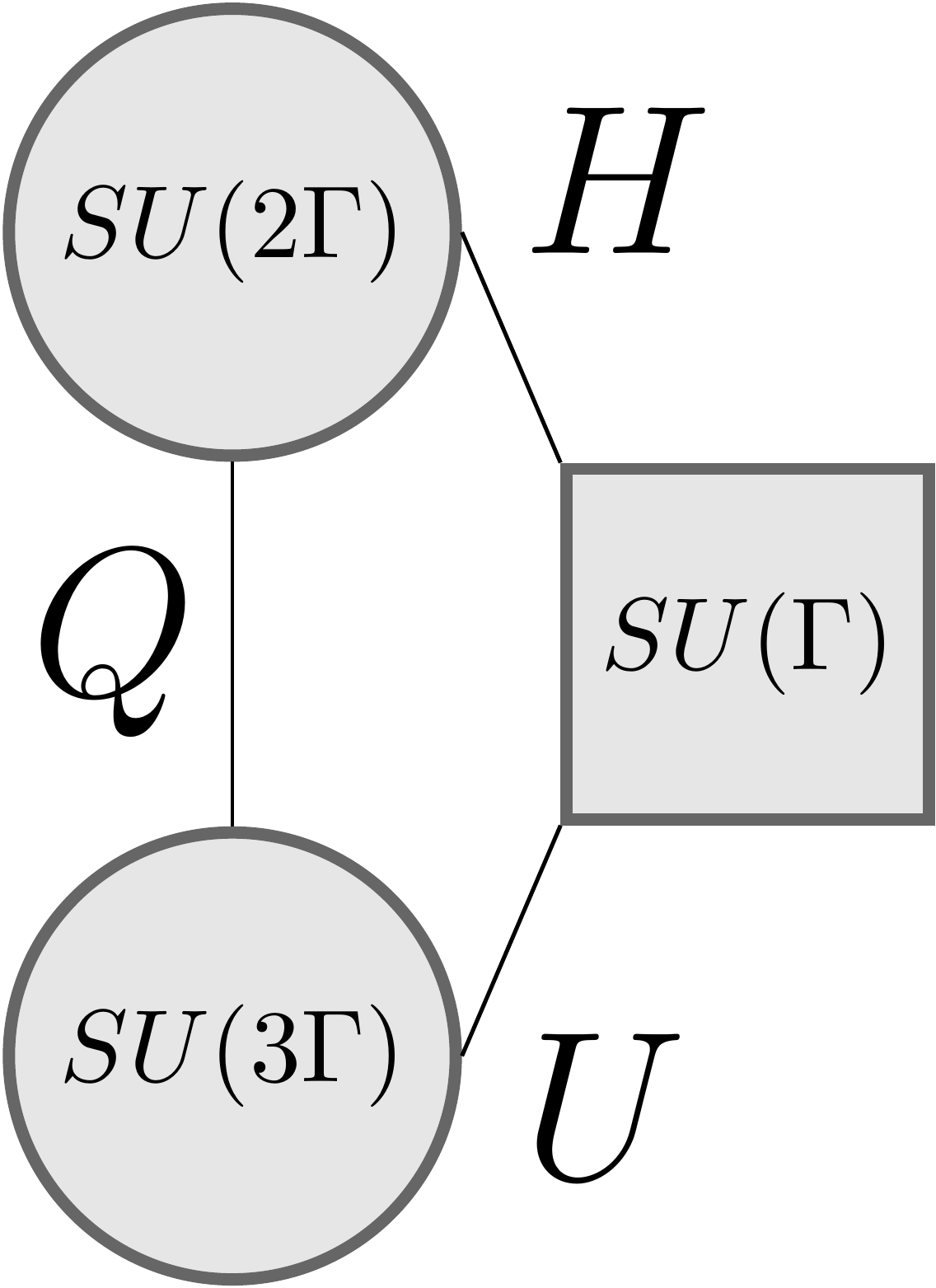} 
}{%
\caption{Quiver diagram of the parent theory. Circular nodes are identified with gauge symmetries and square nodes with flavor symmetries.\label{fig:p_quiver}}%
}
\hspace{15pt}
\capbtabbox{%
\begin{tabular}{c|cc|c}
	&$SU(3\Gamma)$ 		&$SU(2\Gamma)$		&$SU(\Gamma)$		\\	\hline
$H$	&$1$					&$ \square$			& $\overline{\square}$	\\
$Q$	&$ \square$			&$ \overline{\square}$	&$1$					\\
$U$	& $\overline{\square}$	&$1$					&$\square$
\end{tabular}
}{%
  \caption{Matter fields in the parent theory.\label{tab:trans}}%
}
\end{floatrow}
\end{figure}

We can now construct orbifold Higgs models. We begin with a parent theory consisting of a complex scalar, $H$ and fermions, $Q$ and $U$ which transform as bifundamentals of a gauged $SU(2\Gamma) \times SU(3\Gamma)$ and global $SU(\Gamma)$ flavor symmetry. As before, $\Gamma$ will be taken to be the order of the discrete group, $\mcal{G}$, used to construct the daughter theory. The matter content is shown in Table \ref{tab:trans} and a quiver diagram in Figure \ref{fig:p_quiver} .

\begin{figure}[t]
	\includegraphics[scale = .2]{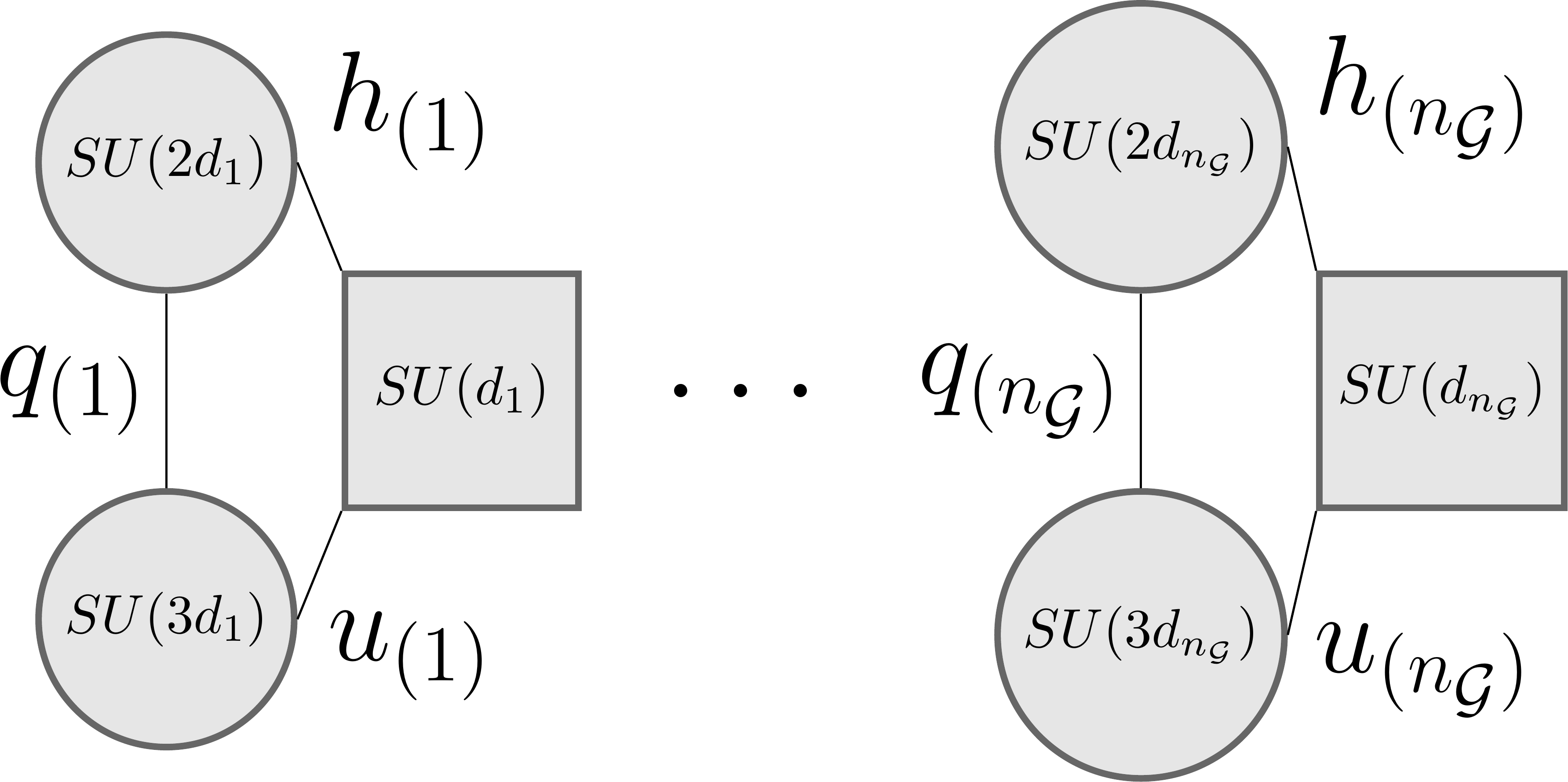}
	\captionsetup{width=.9\linewidth}
	\caption{Quiver diagram of the daughter theory resulting from the orbifold reduction of the parent theory.
	\label{fig:d_quiver}}
\end{figure}

The scalar potential of the parent theory including the Yukawa interactions is given by
\beq
V_P \supset -m^2|H|^2 +\lambda \left(|H|^2\right)^2 +y Q H U .
\eeq
From here we follow the orbifold procedure sketched out above to 
project out the invariant states of the parent theory. The parent theory 
will descend to a daughter theory which can be described by a quiver 
diagram with $n_\mcal{G}$ sets of disconnected nodes, each of which 
resemble the original structure of parent theory as seen in 
Figure \ref{fig:d_quiver}. Each disconnected diagram corresponds to
 a distinct sector charged only under the gauge fields in it's own sector
 \footnote{This true up to $U(1)s$ in the daughter theory which will in 
 general charge multiple sectors. We will address consequences of the 
 residual $U(1)$ factors in section 4.}.  

The potential of the daughter theory takes the form,
\beq
V_d \supset -m^2 \sum\limits_{\alpha=1}^{n_\mcal{G}} |h_\alpha|^2  + \lambda \left(  \sum\limits_{\alpha=1}^{n_\mcal{G}} |h_\alpha|^2 \right)^2 +\sum\limits_{\alpha=1}^{n_\mcal{G}} \frac{y}{\sqrt{d_\alpha}} q_\alpha h_\alpha u_\alpha .
\eeq
The scalar quartic interactions in the daughter theory allows interactions 
between fields in each sector, not unlike in the Twin Higgs. Note the tree 
level scalar potential inherits an accidental $SU(2 \Gamma)$ symmetry. 
There is also a residual discrete symmetry in the scalar sector equivalent 
to the symmetry group leaving the tuple $\{d_1, d_2, ... , d_{n_\G} \}$ 
invariant. These accidental symmetries may however be broken by radiative 
corrections due the gauge and Yukawa interactions. 

Solving for the leading order radiative corrections to the scalar potential we find,
\bea
V^{(1)} &\supset& \frac{\Lambda^2}{16 \pi^2} \left( - 6 y^2 + 3 g_2^2 + (4 \Gamma + 2) \lambda \right)  
			\left( \sum_{\alpha=1}^{n_\mcal{G}} |h^{(\alpha)}|^2 \right) \\
	&-& \frac{3 g_2^2}{64 \pi^2} \left(\sum_{\alpha=1}^{n_{\mcal{G}}}\frac{1}{d_\alpha^2}|h^{(\alpha)}|^2\right)\Lambda^2.
\eea 
Note the corrections in the first line share the accidental $SU(2\Gamma)$ 
symmetry of the tree level potential. One may have naively expected the 
quark yukawas to spoil this accidental symmetry but there is a fortunate 
cancelation of the rescaled couplings with the extra color factors.  It is only 
the gauge interactions at leading order which spoil the accidental $SU(2\Gamma)$ 
symmetry and can contribute to the masses of the would be Goldstones. 

The most simple example of an orbifold Higgs is to take the discrete
 group $\mcal{G} = \mathds{Z}_2$. We would then begin with a parent theory 
 with fields transforming under $SU(6) \times SU(4)$ gauge groups and a 
 $SU(2)$ global symmetry. Upon orbifolding this theory by $\mathds{Z}_2$ 
 the parent theory would descend to a daughter theory with two sectors, each 
 charged under a copy of $SU(3)\times SU(2)$ . This is nothing more but the 
 Twin Higgs! The tree level potential of the daughter theory has the desired 
 accidental $SU(4)$ global symmetry and a discrete symmetry of $\mathds{Z}_2$ 
 which arises as a consequence of the orbifold reduction of the parent theory whereas in the 
 Twin Higgs it was posited as a means to eliminate quadratic divergences.

\section{$S_3$-Orbifold Higgs }
%======================================================%

With the formalism developed we are now equipped to build up the $S_3$ orbifold Higgs model. 
We begin with the potential of the parent theory
\bea
V_P &=& -m^2|H|^2 +\lambda \left(|H|^2\right)^2 +y Q H U
\eea
where the fields transform as bifundamentals under a $SU(18) \otimes SU(12)$ gauge symmetry and a global $SU(6)$
 flavor symmetry. 

\begin{figure}[h]
	\includegraphics[scale = .2]{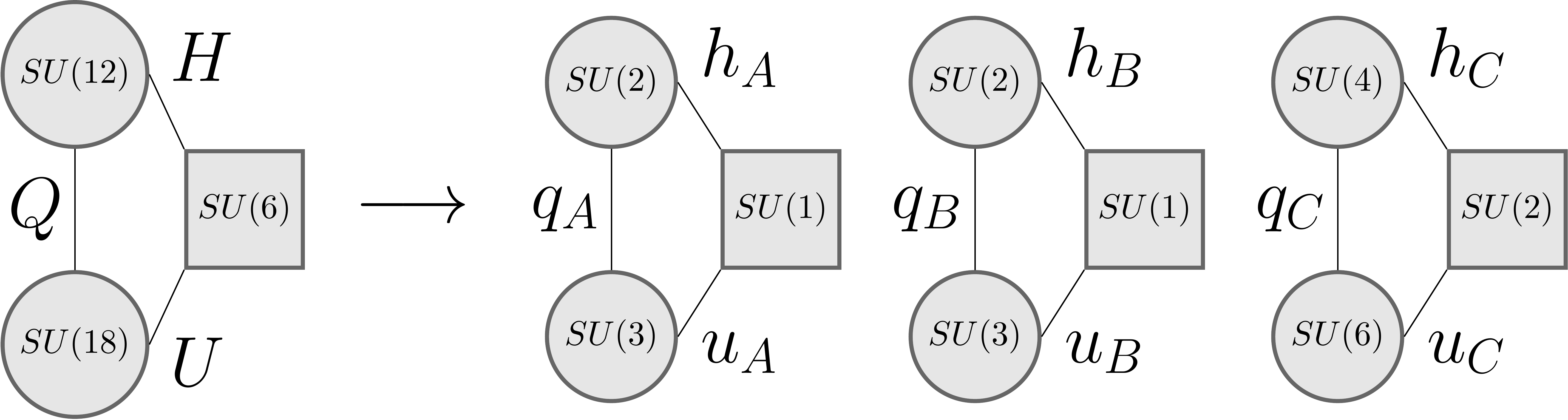} 
	\captionsetup{width=.9\linewidth}
	\caption{Quiver diagram of the parent and daughter theory resulting from the $S_3$-orbifold reduction. The trivial $SU(1)$ nodes
			are drawn only to demonstrate the connection to the parent theory.
	\label{fig:s3quiver}}%
\end{figure}

We will now construct the daughter theory using $\mcal{G} = S_3$ which has 3 
irreducible representations: one dimensional trivial and sign representations, and a single two dimensional 
representation. It follows that we expect three different sectors each charged under its 
own gauge groups, two of which will look standard model like in structure, and a third 
exotic sector with larger gauge groups and a residual flavor symmetry. The quivers of the parent and daughter 
theories are given in Figure \ref{fig:s3quiver}. The invariant combinations of the parent 
fields that survive the orbifold projection and comprise the daughter theory of the 
$S_3$-orbifold Higgs model were worked out and are given in ref. \cite{Craig:2014roa}.

The tree level Higgs potential of the daughter theory is then
\bea
V^{(0)}_d &=& -m^2\left( |h_A|^2 + |h_B|^2 + |h_{C_1}|^2 + |h_{C_2}|^2 \right)\\
    &+& \lambda \left( |h_A|^2 + |h_A|^2 + |h_{C_1}|^2 + |h_{C_2}|^2 \right)^2 \\
	&+& y \bar{q}_A h_A u_A +y \bar{q}_B h_B u_B + \frac{y}{\sqrt{2}} \bar{q}_C h_{C_1} u_{C_1} + \frac{y}{\sqrt{2}} \bar{q}_C h_{C_2} u_{C_2}.
\eea
We use the subscripts $C_1$ and $C_2$ to distinguish the residual $SU(2)$ flavor
 symmetry. Note the factors of $1/\sqrt{2}$ in the c-sector Yukawa interactions. 
 This comes from the rescaling of terms related to the relative dimension of the irreducible representation. 

We now need to include the radiative corrections which will allow us to study the vacuum
 alignment. The dominant contribution to the one-loop effective potential comes from the 
 top loops, 
\bea
V^{(1)}_d &\supset& \frac{3 y^4}{16 \pi^2}
	 \left[  |h_A|^4 \log\left(\frac{\Lambda^2}{y^2 |h_A|^2} \right)
	       + |h_B|^4 \log\left(\frac{\Lambda^2}{y^2 |h_B|^2} \right) \right. \\
   &+& \left. \frac{1}{2}|h_{C_1}|^4 \log\left(\frac{\Lambda^2}{\frac{y^2}{2} |h_{C_1}|^2} \right) 
   	      +  \frac{1}{2}|h_{C_2}|^4 \log\left(\frac{\Lambda^2}{\frac{y^2}{2} |h_{C_2}|^2} \right) \right].
\eea
Adding this contribution to the tree level scalar potential we find that 
$|\langle h_A\rangle|^2 = |\langle h_B\rangle|^2 =\frac{1}{2} |\langle h_{C_1}\rangle|^2 =
 \frac{1}{2} |\langle h_{C_2}\rangle|^2 = \frac{1}{2} \frac{m^2}{6 \lambda + \delta} \equiv  \frac{1}{12}f^2 $.
  At this point none of sectors can be identified with the SM-like sector due to the fact that the weak 
  scales aren't adequately separated causing this Higgs to be not well aligned with the SM sector.     

To remedy this we add a soft term of the form,
\bea
V_{soft} &=& \rho^2 \left(   |h_A|^2 - \frac{1}{5} |h_B|^2 - \frac{1}{5}|h_{C_1}|^2 - \frac{1}{5} |h_{C_2}|^2 \right) \\
		&+& \sigma^2 \left(h_{C_1} - h_{C_2} \right)^\dagger \left(h_{C_1} -h_{C_2} \right)
\eea 
which will allow us to identify the A-sector with the SM-like sector.
The first piece is used to break the residual $S_2$ symmetry of the daughter theory. 
The specific form is chosen only to simplify future expressions for the vevs and masses. 
A more general expression would alter the alignment between the $B$ and $C$-sectors,
but this plays a modest role in determining the phenomenology of the SM-like Higgs.
 The second term is added to allow the would be
Goldstones in the $C$-sector to acquire mass.

The addition of a soft term makes it difficult to gain analytic expressions for these quantities
 so we introduce the following approximation. We approximate  
\beq
\frac{3 y^4}{16 \pi^2} \log\left(\frac{\Lambda^2}{\frac{y^2}{d_\alpha} |h_\alpha|^2} \right)
\approx \frac{3 y^4}{16 \pi^2} \log\left(\frac{\Lambda^2}{y^2 |\langle h_A\rangle |^2} \right)\equiv \delta, \,\,\,\,\,\mbox{for }\,\, \alpha = A,B,C_1,C_2.
\eeq
This does remove the dynamics of the fields within the logarithm but those have a much 
smaller effect compared to the dynamics in the multiplicative factor of $|h_\alpha|^4$ in determining the
vacuum alignment. The approximation is reasonable for $f \lesssim \mbox{few} \times |\langle h_A \rangle |$. 

Working from the approximate potential of the daughter theory,
\bea
V_d  &\supset&  -m^2 (|h_A|^2 + |h_B|^2 + |h_{C_1}|^2 + |h_{C_2}|^2) \\
	&+& \lambda \left( |h_A|^2 + |h_B|^2 + |h_{C_1}|^2 + |h_{C_2}|^2 \right) \\
	&+& \rho^2 \left(   |h_A|^2 - \frac{1}{5} |h_B|^2 - \frac{1}{5}|h_{C_1}|^2 - \frac{1}{5} |h_{C_2}|^2 \right) \\
	&+& \sigma^2 \left(h_{C_1} - h_{C_2} \right)^\dagger \left(h_{C_1} -h_{C_2} \right) \\
	&+& \delta \left(  |h_A|^2 + |h_B|^2 + \frac{1}{2}|h_{C_1}|^2 + \frac{1}{2}|h_{C_2}|^2  \right)
\eea
we find the following expressions for the vevs,
\beq
	\begin{array}{ccccc}
	v^2 \equiv 2 |\langle h_A\rangle|^2 = \frac{m^2}{6\lambda +\delta} - \frac{\rho^2}{\delta}, &	&v^2_B =  \frac{m^2}{6\lambda +\delta} + \frac{\rho^2}{5\delta}, 	&	&	v^2_{C_1}  =v^2_{C_2} =  \frac{2m^2}{6\lambda +\delta} + \frac{2\rho^2}{5\delta}.
	\end{array}
\eeq
Tuning $\frac{\rho^2}{\delta}$ against $\frac{m^2}{6\lambda +\delta}$ allows us to achieve
 a vacuum alignment that is consistent with the $A-$sector being associated with the SM
  like sector in the theory. This corresponds to a tree-level tuning on the order of $6v^2/f^2$.

Upon diagonalization of the mass matrix  we find the SM-like Higgs, 
$h \approx \cos(v/f) \phi^3_A -\frac{1}{\sqrt{5}}\sin(v/f) \left( \phi^3_B + \sqrt{2}\phi^7_{C_1}   
+ \sqrt{2} \phi^7_{C_2}  \right)$ where the $\phi^i_\alpha$ fields are the components $h_\alpha$
 in the hermitian basis given in Eq. (\ref{eq:hermitianfields}) of the Appendix. The corresponding
  mass of $h$ is found to be $m_h^2 \approx \frac{12}{5}\delta f^2 \sin^2 \left( \frac{v}{f}\right)$. 
  The remaining mass eigenstates are listed in the Appendix.

\subsection{U(1) Daughter Gauge Fields}
%======================================================%

Up to now we have set aside the residual $U(1)$ factors of the daughter theory as 
they play little importance in the determining the vacuum alignment. We'll now take 
a moment to discuss some possibilities for handling these extra fields.  A simple option 
would be to set them aside or lift the $U(1)$ fields via the Stueckelberg mechanism\cite{Stueckelberg:1900zz,Ruegg:2003ps}, leaving behind no massless gauge fields that 
interact with multiple sectors. Hypercharge assignments, at least the SM sector, can then be added in that 
would break the orbifold correspondence to the mother theory and will contribute
 additional radiative corrections to the Higgs effective potential. This will be the path we take in analyzing the collider signatures of the model 
 in  section \ref{sec:pheno}.

Another interesting possibility is to take a linear combination of the $U(1)$s and identify
 it with the hypercharge generator and lifting the remaining $U(1)$s through the 
 Stueckelberg mechanism. In this case the hypercharge generator will charge the SM and  $C$-sector
 which places additional constraints from precision electroweak measurements and charged dark
 matter searches on this scenario. 

\section{Phenomenology} \label{sec:pheno}
%======================================================%

In this section we apply a similar analysis to\cite{Burdman:2014zta}, whereby we 
calculate the modifications to Higgs production cross sections and branching fractions.
We will then compare our results with those predicted by the Mirror Twin Higgs model. Lastly,
we will discuss the tuning and naturalness of the model.

We expect the production cross sections and decay widths 
to SM particles of the $125\mbox{ GeV}$ Higgs, $h$, to suppressed by a multiplicative 
factor of $\cos^2(v/f)$ giving us,
\bea
\sigma(pp \to h) & = &\cos^2 \left(v/f  \right) \sigma(pp \to h_{SM}) \\
\Gamma(h \to SM_i) & = & \cos^2 \left(v/f  \right) \Gamma(h_{SM} \to SM_i)
\eea
where the subscript, $i$,  denotes some particular final state. For $f = \mbox{few }\times v$, this 
is consistent with the SM prediction. 

The decay widths of $h$ to the hidden sector states should be suppressed by a factor 
of $\sin^2(v/f)$ from the Higgs alignment but should also be accompanied by another 
multiplicative factor stemming from kinematical effects. It is convenient to define the dimensionless 
quantities,
\beq
r_B \equiv \frac{\Gamma(h \to \mbox{$B$-sector})}{\Gamma(h_{SM}) \frac{1}{5}\sin^2 \left(v/f  \right)}
\hspace{10mm}
\mbox{and}
\hspace{10mm}
r_C \equiv \frac{\Gamma(h \to \mbox{$C$-sector})}{\Gamma(h_{SM}) \frac{2}{5}\sin^2 \left(v/f  \right)}
\eeq
which will allow us to simply cast the total width of the Higgs as,
\beq
\Gamma(h) =    \Gamma(h_{SM}) \left[ \cos^2 \left(v/f  \right) +\frac{1}{5}\sin^2(v/f)\left(r_B   
			+ 2 r_C  \right)  \right].
\eeq
Using the above relations we can write signal strength for Higgs decays into SM particles as
\beq
\frac{\sigma(pp \to h) BR(h \to SM_i)}{\sigma(pp \to h_{SM}) BR(h_{SM} \to SM_i)} = 
\frac{\cos ^2\left(v/f \right)}{1+ \frac{1}{5} (r_B+2 r_C) \tan ^2\left(v/f \right)},
\eeq 
where $r_{B/C}$	 now need to be determined. 
		
\begin{table}[t]
\centering
\captionsetup{width=.9\linewidth}
\caption{	A summary of common SM Higgs boson decays\cite{Djouadi:2005gi} that we'll consider in our analysis 
		of the SM-like Higgs decays into b and c-sector states. The  $\delta_V$, $R_T$, and 
		$A_X$ functions are defined in the Appendix.  }
\label{tab:h_decays}
\begin{tabular}{|c|l|}
\hline
\multicolumn{2}{|c|}	{Standard Model Higgs Decays}           \\ \hline\hline
\multicolumn{2}{|l|}	{ \hspace{5mm}$\Gamma(h \to \overline{f}f) = \frac{N_c}{16\pi}m_h\lambda_f^2\left(1-4\frac{m_f^2}{m_h^2} \right)^{3/2}$\hspace{5mm} }  \\ \hline
\multicolumn{2}{|l|}	{ \hspace{5mm}$\Gamma(h\to VV^{\ast}) = \frac{3 m_h}{32\pi^3}\frac{m_V^4}{v^4_{\text{EW}}} \delta_VR_T\left(\frac{m_V^2}{m_H^2}\right)$\hspace{5mm} }\\ \hline
\multicolumn{2}{|l|}	{ \hspace{5mm}$\Gamma(h \to g g) = \frac{\alpha_s^2 m_h^3}{72\pi^3v^2}\left|\frac34 \sum\limits_q A_F\left(\frac{4m_q^2}{m_h^2} \right) \right|^2$\hspace{5mm}  } \\ \hline
\multicolumn{2}{|l|}	{ \hspace{5mm}$\Gamma(h \to \gamma \gamma) = \frac14 \frac{e^4 m_h^3}{(4\pi)^5v^2}\left|\ \sum\limits_q A_F\left(\frac{4m_q^2}{m_h^2} \right)  + A_V\left(\frac{4m_V^2}{m_h^2} \right) \right|^2$\hspace{5mm} }\\ \hline
\end{tabular}
\end{table}

	Before proceeding directly to the calculation it is worth recalling the leading order partial widths for 
SM Higgs to fermions, vector bosons, gluons, and photons which are summarized in Table \ref{tab:h_decays}. 
The expression for $r_B$ follows directly from\cite{Burdman:2014zta} and is given by,
\begin{align}
r_B=&\sum_j \mbox{BR}(h\to f_j\overline{f}_j) \left[ \frac{\displaystyle 1-4\frac{m^2_{f_j}}{m_h^2} \frac{v_B^2}{v^2}}{\displaystyle 1-4\frac{m^2_{f_j}}{m_h^2}}\right]^{3/2} +
\sum_j  \frac{\delta_{V_j}(\theta_W \to 0)}{\delta_{V_j}} \mbox{BR}(h\to V_jV^{\ast}_j)\frac{\displaystyle R_T\left(\frac{m_{V_{j}}^2}{m_h^2}\frac{v_B^2}{v^2}\right)}{\displaystyle R_T\left( \frac{m_{V_{j}}^2}{m_h^2}\right)}\nonumber\\
&+\mbox{BR}(h\to gg)\frac{\displaystyle \left|A_F\left( \frac{4m_{t}^2}{m_h^2}\frac{v_B^2}{v^2}\right)\right|^2}{\displaystyle \left|A_F\left( \frac{4m_t^2}{m_h^2}\right) \right|^2}.
\end{align} 
The Weinberg angle is set to zero since we've excluded the hypercharge in the hidden sectors.

The expression for $r_C$ is slightly complicated by the scaled couplings, and larger color factors. The
massive gauge bosons kinematically forbid decays of $h \to V_c^\ast V_c$ for the ranges of the order 
breaking parameter $f$ we consider here. However, loop level decays to the 8 massless gauge bosons 
now contribute to the width
\footnote{Depending on sign of the beta function for the $SU(6)$ color group 
this sector may confine and Higgs the remaining $SU(3)$ subgroup. We'll proceed 
assuming gauge bosons of  the $SU(3)$ subgroup remain massless thus placing more conservative
bounds on the model.  }. 
We can modify the SM Higgs decay width to two photons to express the decay width to massless gauge bosons and express $r_C$ as, 
\begin{align}
r_C=&2 \left( \sum_{j}\mbox{BR}(h\to f_j\overline{f}_j) 
\left[ \frac{\displaystyle 1-4\frac{m^2_{f_j}}{2m_h^2} \frac{v_{C_1}^2}{v^2}}{\displaystyle 1-4\frac{m^2_{f_j}}{m_h^2}}\right]^{3/2} \right. +
\frac{35}{32} \mbox{BR}(h\to gg)
\frac{\displaystyle \left|A_F\left(  \frac{4m_{t}^2}{m_h^2}\frac{v_{C_1}^2}{2 v^2}\right) \right|^2}{\displaystyle \left|A_F\left( \frac{4m_t^2}{m_h^2}\right) \right|^2} \nonumber \\
&+
 2 \cdot \frac{v^2}{v_{C_1}^2+v_{C_2}^2}\frac{g^4}{e^4} 
\left. \mbox{BR}(h\to \gamma \gamma)
\frac{\displaystyle \left|A_W\left( \frac{4m_{W}^2}{m_h^2}\frac{v_{C_1}^2+v_{C_2}^2}{2v^2}\right) \right|^2}{\displaystyle \left|A_W\left( \frac{4m_t^2}{m_h^2}\right) \right|^2}  \right)
\end{align} 

	We are now left to calculate $r_{B/C}$ to attain the signal strength of the Higgs into SM particles and 
the branching ratio of Higgs to hidden sector states. We'll assume a 3 generation model of quarks and leptons.
This assumption is problematic when considering the thermal history of the universe where copies of light generations
 could alter $N_{eff}$. However adding in the down type quarks and extra generations predicts a larger branching 
 fraction of Higgs to hidden states, thus providing more conservative estimates for the decay rates to hidden sector states.

\begin{figure}[t]
\includegraphics[scale = .42]{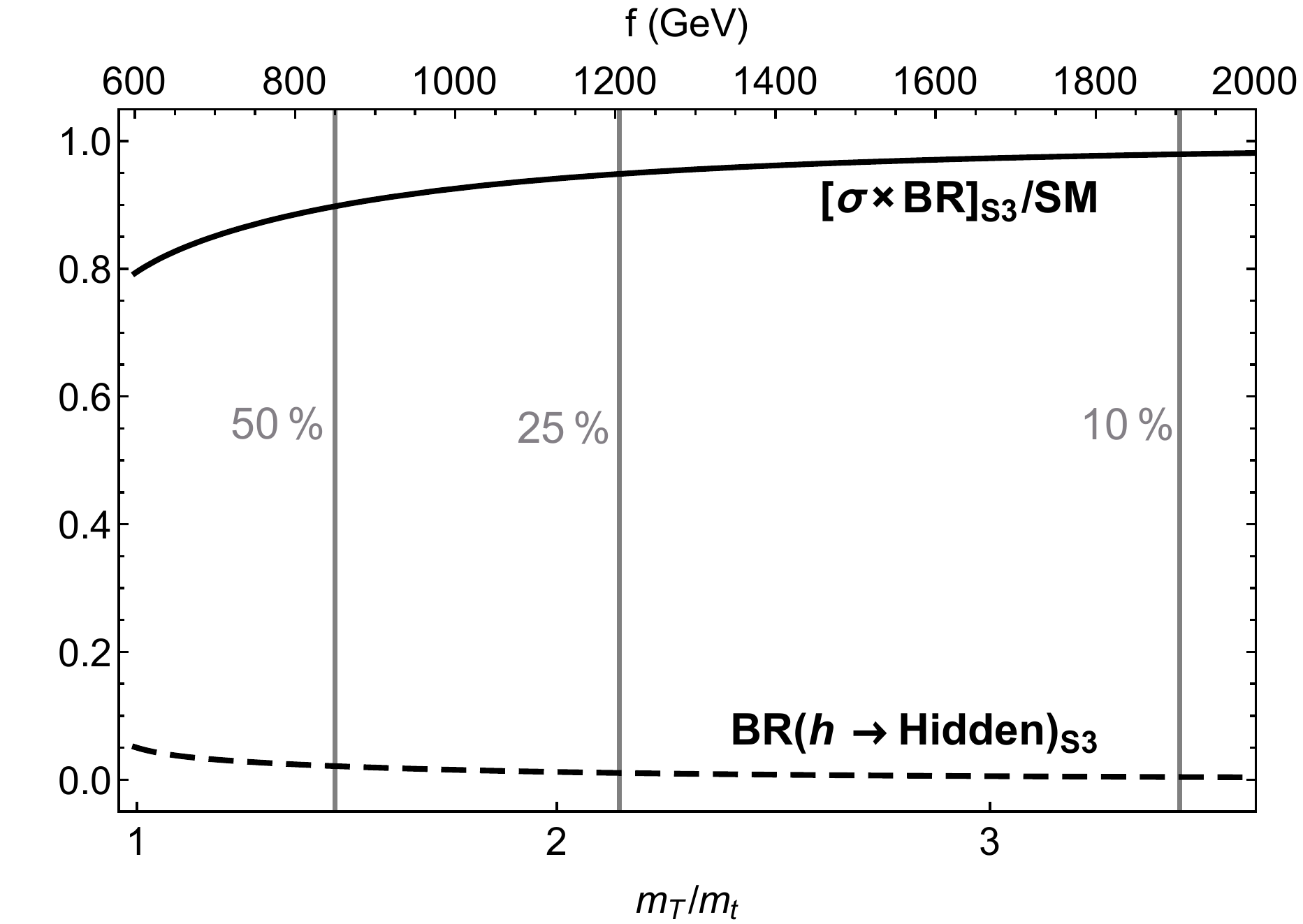}
\includegraphics[scale = .43]{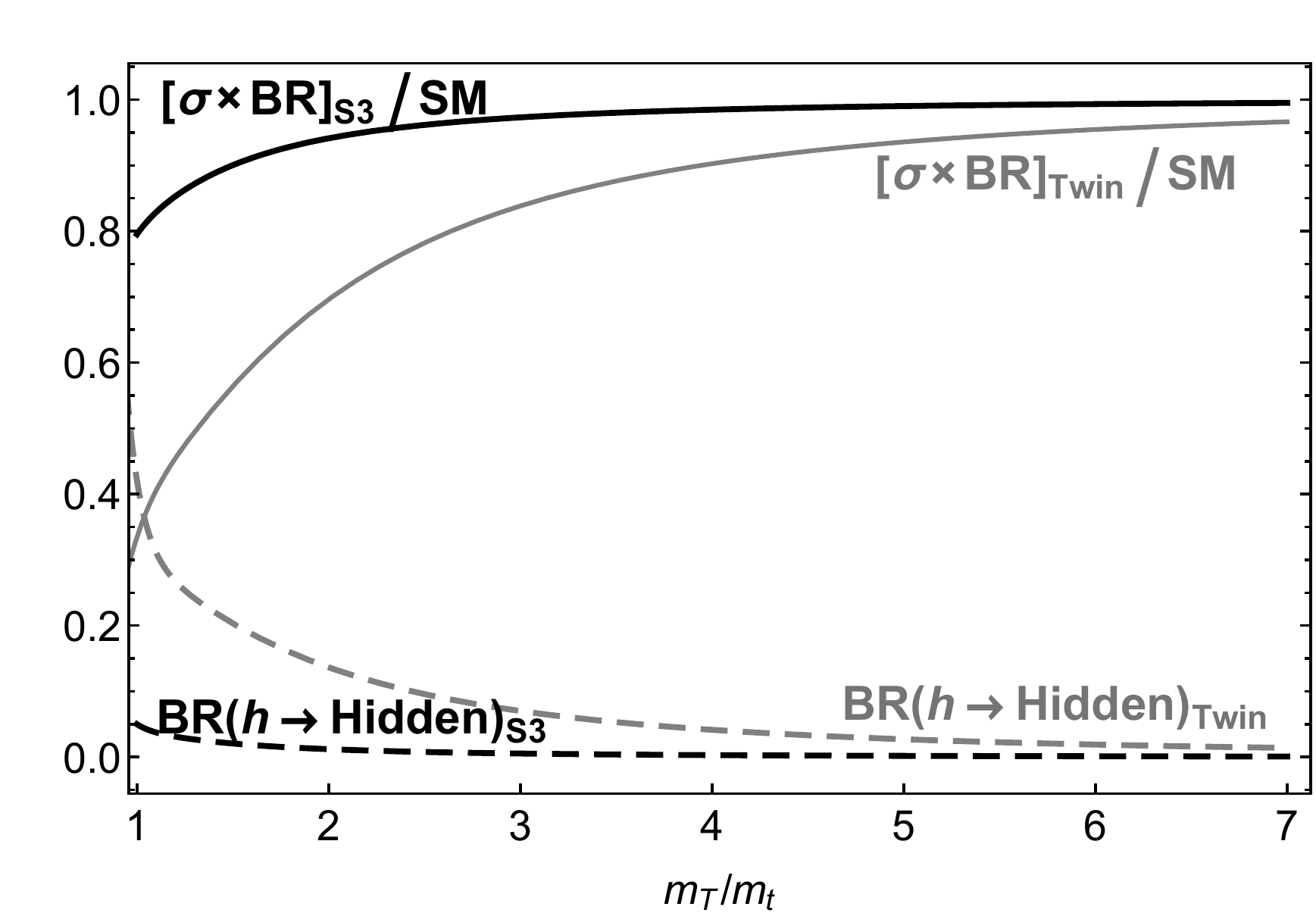}
\captionsetup{width=.9\linewidth}
\caption{(Left) Plot of the signal strength of producing $h$ and it directly decaying to SM particles and the branching ratio for decays $h$ decays to 
		A and B sector particles. The plot is given as a function of both the order breaking parameter $f$ and the ratio of the top partner masses divided 		by the SM top mass. (Right) Comparison plot of the signal strength and branching fraction of hidden sector $h$ decays plotted as a function of the 		top partner mass ratios.}
\label{fig:plots}
\end{figure}

In Fig. \ref{fig:plots} we present plots for the signal strength of the 
SM-like Higgs and it's branching fraction to hidden states. We give a plot 
comparing these results to those given in ref. \cite{Burdman:2014zta} for the Twin Higgs. Though the
behavior is very similar we note that the $S_3$-orbifold Higgs model approaches 
the SM result faster as a function of top partner mass. This stems from the fact that 
vev is now shared across three sectors allowing for lighter partner states
for a given $SU(6)$ breaking order parameter, $f$, as compared to the Twin Higgs partner states. 

	Let's now consider the level of tuning occurring in model. In Eq. 3.9 we found the 
leading order radiative corrections of the scalar potential that break
the accidental $SU(2 \Gamma)$ symmetry of the tree level potential in a general 
orbifold Higgs model. In the case of the $S_3$-orbifold theory at hand this corresponds to 
\beq
\delta m^2 \approx \frac{3 g_2^2}{64 \pi^2} \Lambda^2 \left(1 -  \frac{1}{2^2}   \right).
\eeq
Using
\beq
\Delta_m = \left|  \frac{2 \delta m^2}{m_h^2} \right|^{-1}
\eeq	
as an estimate of our tuning, corresponds to a 50\%, 25\%, and 10\% level tuning
at cutoff scales of 3.3 TeV, 4.7 TeV , and 7.5 TeV respectively.

	As mentioned in Section \ref{sec:pheno} in order to associate the $A$-sector with the
SM-like sector we needed to tune $\frac{\rho^2}{\delta}$ against $\frac{m^2}{6\lambda +\delta}$
which resulted in a modest tuning of order $6v^2/f^2$. A tree level tuning of 50\%, 25\%, and 10\% corresponds to
 an $SU(12)$ breaking order parameter of $f \approx $ 0.85 TeV, 1.2 TeV, and 1.9 TeV respectively, or equivalently 
in terms of the top partner mass of $m_T \approx$ 1.48$m_t$, 2.14$m_t$, and 3.43 $m_t$. We overlay the plot in 
Figure \ref{fig:plots} with lines indicating these tree level tunings.

\section{Results and Future Prospects} \label{sec:c&o}
	
	The $S_3$-orbifold Higgs can easily accommodate the SM
without facing any major tuning for cutoff scales approaching $8$ TeV. A 10\%
tree level tuning is sufficient to give the signal strength the SM Higgs within a couple percent. 
Though the nature of the model may seem complicated with three sectors which can only communicate
through the Higgs portal, the Higgs phenomenology is only dependent on two additional parameter to 
the SM, the $SU(12)$ breaking order parameter $f$ and the soft term $\rho$. This makes
the testability of model in principle no more complicated than Twin Higgs. 

	The LHC has greater sensitivity in measuring signals from SM decays of the Higgs compared to invisible decays. This
makes searching for deviations in SM Higgs decay channels favorable for testing the model.
At an integrated luminosity of $3000\mbox{ fb}^{-1}$ the LHC will be able to probe Higgs signal strengths 
in the $WW$, $ZZ$, and $\gamma \gamma$ channels down to the 5\% level\cite{Dawson:2013bba}.
If a suppression in the signal strengths of more than 5\%  is measured, the model 
will be pushed into the region of parameter space where with top partner masses $m_T \lesssim 2 m_t$.
 This makes it difficult for the LHC to strongly disfavor the 
$S_3$-orbifold Higgs as a natural model. The increased Higgs production of a 100 TeV collider however may 
provide a way of testing the naturalness of the model.

	There is also the possibility for more exotic collider signatures in the form of Higgs decays with displaced vertices. It is possible for 
the SM-like Higgs to decay into $B$ and $C$-sector states which may decay back into SM states giving
rise to so-called ``hidden valley" signatures\cite{Strassler:2006im,Strassler:2006ri,Han:2007ae}. 
These signatures were studied in the context of the Fraternal Twin Higgs\cite{Craig:2015pha}. 
The phenomenology in the $S_3$-orbifold Higgs model should be qualitatively similar. A thorough
comparison would require a more detailed study of the hidden sectors and mass scales of the glueballs produced
in each sector, including those that may be produced by the unbroken $SU(3)$ subgroup of the broken $SU(4)$ weak gauge
group in the $C$-sector.
 
 	An interesting feature of the model is that for relatively light top partners, in comparison 
to those in the Twin Higgs, there is still a large suppression of Higgs decays to hidden sector states. This
 is a general feature of orbifold Higgs models where the orbifold projection produces three or more sectors.
With such a modest difference in the masses of fermion partners 
it may be interesting to study if any of the matter in the hidden sectors could serve as a
stable dark matter candidate. The possibility of $C-$sector having multiple confining 
gauge groups in the theory may also provide additional stability against the states decaying into SM states. 
 There have already been a number of dark matter and cosmology studies involving the Twin Higgs\cite{Craig:2016lyx, Chacko:2016hvu, Barbieri:2016zxn,
Prilepina:2016rlq,Farina:2015uea,Farina:2016ndq,Garcia:2015toa,Craig:2015xla,Garcia:2015loa,Csaki:2017spo}
 which may serve as an avenue for future work involving the $S_3$-orbifold Higgs model.

\section{Acknowledgments}
Many thanks to Marc Sher, Christopher Carone, and Joshua Erlich for thoughtful 
discussions. This work was supported by the NSF under Grant PHY-1519644. 

\appendix
\newpage
\section{Appendix}
%======================================================%

Scalar multiplets of the daughter theory in the Hermitian basis. 

\beq
\begin{array}{ccc}
h_A = \frac{1}{\sqrt{2}}
\left(
	\begin{array}{c}
 	\phi_A^1 + i \phi_A^2 \\
  	(v_A+ \phi_A^3) + i \phi_A^4
	\end{array}  
 \right) & 
\hspace{2mm} &
h_B = \frac{1}{\sqrt{2}}
\left(
	\begin{array}{c}
 	\phi_B^1 + i \phi_B^2 \\
  	(v_B+\phi_B^3) + i \phi_B^4
	\end{array}  
  \right) \\
  & & \\
h_{C_1} = \frac{1}{\sqrt{2}}
	\left(
	\begin{array}{c}
 	\phi_{C_1}^1 + i \phi_{C_1}^2 \\
 	\phi_{C_1}^3 + i \phi_{C_1}^4 \\
 	\phi_{C_1}^5 + i \phi_{C_1}^6 \\
  	(v_{C_1}+\phi_{C_1}^7) + i \phi_{C_1}^8   
\end{array}  
\right) &
\hspace{2mm} &
h_{C_2} = \frac{1}{\sqrt{2}}
	\left(
	\begin{array}{c}
 	\phi_{C_2}^1 + i \phi_{C_2}^2 \\
 	\phi_{C_2}^3 + i \phi_{C_2}^4 \\
 	\phi_{C_2}^5 + i \phi_{C_2}^6 \\
  	(v_{C_2}+\phi_{C_2}^7) + i \phi_{C_2}^8   
\end{array}  
\right) 
\end{array}
\label{eq:hermitianfields}
\eeq

The scalar mass eigenstates given in terms of the fields of the Hermitian basis with $\theta = v/f$ and $c_\theta \equiv \cos \theta$ $s_\theta \equiv \sin \theta$.

\beq
\left(
\begin{array}{c}
h \\
H_1 \\
H_2 \\ 
H_{radial}
\end{array}
\right)
=
\left(
\begin{array}{cccc}
c_\theta  	&-\frac{1}{\sqrt{5}}s_\theta	&-\frac{\sqrt{2}}{\sqrt{5}}s_\theta  &-\frac{\sqrt{2}}{\sqrt{5}}s_\theta    \\
0  		&\frac{2}{\sqrt{5}}	& \frac{-1}{\sqrt{10}}  &\frac{-1}{\sqrt{10}}   \\
0  		&0	&   \frac{1}{\sqrt{2}}& \frac{-1}{\sqrt{2}}  \\
s_\theta  		&\frac{1}{\sqrt{5}}c_\theta	&\frac{\sqrt{2}}{\sqrt{5}}c_\theta   &\frac{\sqrt{2}}{\sqrt{5}} c_\theta 
\end{array}
\right)
\left(
\begin{array}{c}
\phi_A^3 \\
\phi_B^{3} \\
\phi_{C_1}^{7} \\ 
\phi_{C_2}^{7}
\end{array}
\right)
\eeq

\beq
\begin{array}{ccc}
H_3=\frac{1}{\sqrt{2}} \left( \phi_{C_1}^1 - \phi_{C_2}^1  \right) & 
H_4=\frac{1}{\sqrt{2}} \left( \phi_{C_1}^2 - \phi_{C_2}^2  \right) &
H_5=\frac{1}{\sqrt{2}} \left( \phi_{C_1}^3 - \phi_{C_2}^3  \right)  \\
H_6=\frac{1}{\sqrt{2}} \left( \phi_{C_1}^4 - \phi_{C_2}^4  \right) &
H_7=\frac{1}{\sqrt{2}} \left( \phi_{C_1}^5 - \phi_{C_2}^5  \right) &
H_8=\frac{1}{\sqrt{2}} \left( \phi_{C_1}^6 - \phi_{C_2}^6  \right) \\
&H_9=\frac{1}{\sqrt{2}} \left( \phi_{C_1}^8 - \phi_{C_2}^8  \right) & 
\end{array}
\eeq

Below we list the corresponding masses for the mass eigenstate given above.

\bea
m^2_{H_1} & \approx & \frac{2}{5}\delta f^2 \cos^2 \left( \frac{v}{f}\right) \\
m^2_{H_2} & \approx & \frac{2}{5}\delta f^2 \cos^2 \left( \frac{v}{f}\right) +2\sigma^2 \\
m^2_{H_{3-9}} &\approx& 2\sigma^2 \\
m^2_{H_{radial}} & \approx & 2 \lambda f^2
\eea

Functions appearing in the Higgs partial decay widths.
\bea
A_V(x) &=&-x^2\left[\frac{2}{x^2}+\frac{3}{x}+3\left(\frac{2}{x}-1 \right)\arcsin^2\left(\frac{1}{\sqrt{x}} \right) \right] \\
A_F(x) &=& 2x^2\left[\frac{1}{x}+\left(\frac{1}{x}-1 \right)\arcsin^2\left(\frac{1}{\sqrt{x}} \right) \right] \\
\delta_W &=& 1 \\
\delta_Z &=& \frac{7}{12} -\frac{10}{9} \sin^2\theta_W + \frac{40}{9} \sin^4\theta_W
\eea
\begin{align}
R_T(x)=&\frac{3(1-8x+20x^2)}{\sqrt{4x-1}}\cos^{-1}\left(\frac{3x-1}{2x^{3/2}}\right)- \frac{1-x}{2x}(2-13x+47x^2)\nonumber\\
&-\frac32(1-6x+4x^2)\ln x\, 
\end{align}

\newpage

%%%%%%%%%%%%%%%%%%%%%%%%%%%%%%%%%%%%%%%%%%%%%%%%%%%%%%%%%%%
%%%%%%%%%%%%%%%%%%%%%%%%%%%%%%%%%%%%%%%%%%%%%%%%%%%%%%%%%%%

\end{document}